\documentclass[preprint,showpacs,showkeys,preprintnumbers,amsmath,amssymb]{revtex4}

\usepackage{graphicx}
\usepackage{dcolumn}
\usepackage{bm}
\usepackage{hyperref}
\hypersetup{colorlinks=false}

\begin{document}


\title{Towards a three dimensional solution for 3N bound states \\ with 3N forces}

 \author{M.~R. Hadizadeh}
\email{hadizade@khayam.ut.ac.ir}
\author{S. Bayegan}%
 \email{bayegan@khayam.ut.ac.ir}

\affiliation{ Department of Physics, University of Tehran, P.O.Box
14395-547, Tehran, Iran }%

\date{\today}

\begin{abstract}
After a brief discussion about the necessity of using the 3D
approach, we present the non PW formalism for 3N bound state with
the inclusion of 3N force (3NF). As an example the evaluation of 3NF
matrix elements, which appear in the obtained coupled three
dimensional integral equations, for $2\pi$-exchange Tucson-Melbourne
3NF show how would be this formalism efficient and less cumbersome
in comparison with the PW formalism.
\end{abstract}

\maketitle

\section{Why do we use 3D approach instead of PW approach?}

The answer to this main question indicates the motivation for using
this approach. Few-body calculations are traditionally carried out
by solving the relevant equations in a PW basis which after
truncation they lead to coupled equations on angular momentum
quantum numbers. A few PWs often provide qualitative insight, but
modern calculations need many different spin, isospin and angular
momentum combinations. It is clear that in PW approach one should
sum all PWs to infinite order, but in practice one truncates the sum
to a finite angular momentum number which is dependent to the energy
that one is working. It means that in higher energies one will need
more PWs to obtain a convergence. In contrast to the traditional PW
representation, the novel 3D approach replaces the discrete angular
momentum quantum numbers with continuous angle variables and
consequently it considers automatically all PW components to
infinite order \cite{Fachruddin-PRC62}-\cite{Bayegan-PTP}. So the
number of equations in 3D formalism is energy independent, also this
formalism avoids the very involved angular momentum algebra
occurring for the permutations, transformations and especially for
the 3NFs.

\section{3D representation of Faddeev equations with 3NF} \label{sec:FC_with_3NF}
We have recently applied the 3D approach to the 3N bound state,
where the Faddeev equations with NN interactions are successfully
solved with Bonn-B potential \cite{Bayegan-PRC77}. In this article
we extend this formalism by considering the 3NF, which helps us to
reach to the full solution of the 3N bound state in a
straightforward manner. The 3N bound state in the presence of the
3NF is described by the Faddeev equation:
\begin{eqnarray}
\label{eq.FC} | \psi \rangle &=& G_{0}tP |\psi\rangle +(1+G_{0}t)
G_{0} V_{123}^{(3)} |\Psi\rangle,
\end{eqnarray}
where the quantity $V_{123}^{(3)}$ defines a part of the 3NF which
is symmetric under the exchange of the particles $1$ and $2$. %
In order to solve the Eq. (\ref{eq.FC}) in momentum space we
introduce the 3N basis states in a 3D formalism as:
\begin{eqnarray}
  |\, {\bf p} \, {\bf q} \,\,  \alpha  \, \rangle \equiv |
 {\bf p} \, {\bf q} \,\, \alpha_{S} \,\, \alpha_{T}  \,  \rangle
 &\equiv& |
 {\bf p} \, {\bf q} \,\, (s_{12} \,\,
 \frac{1}{2}) S \, M_{S} \, \, (t_{12} \,\,
 \frac{1}{2}) T \, M_{T} \, \rangle.
 \label{eq.basis}
\end{eqnarray}
Evaluation of the transition and permutation operators need to the
free 3N basis states $|\, {\bf p}\,{\bf q} \,\, \gamma \, \rangle$,
where the spin-isospin parts $\gamma$ are given as: $|\gamma \,
\rangle \equiv| \gamma_{S} \,\, \gamma_{T} \,
 \rangle \equiv | m_{s_{1}} \, m_{s_{2}} \,  m_{s_{3}} \,\, m_{t_{1}} \, m_{t_{2}} \,
 m_{t_{3}} \, \rangle.  $
To this aim by changing the 3N basis states $| \alpha \, \rangle$ to
the free 3N basis states $| \gamma \, \rangle$ one needs to
calculate the usual Clebsch-Gordan coefficients $\langle \, \gamma |
\alpha \, \rangle = g_{\gamma \alpha } \equiv g_{\gamma \alpha }^{S}
\, g_{\gamma \alpha }^{T}$, see appendix (A) of Ref.
\cite{Bayegan-PRC77}. 
The evaluation of Faddeev equation with the inclusion of 3NF will be
exactly the same as Eq. (19) of Ref. \cite{Bayegan-PRC77} except
that an extra term with $V_{123}^{(3)}$. This is
\begin{eqnarray}  \label{Eq.3NF-term}
 \langle \, {\bf p}\, {\bf q}\, \alpha | (1+G_{0}t) G_{0} V_{123}^{(3)} |\Psi\rangle
   = \frac{1}{{E-\frac{p^{2}}{m}
-\frac{3q^{2}}{4m}}}
 \biggl\{ \langle \, {\bf p}\, {\bf q}\, \alpha |
V_{123}^{(3)} |\Psi\rangle + \langle \, {\bf p}\, {\bf q}\, \alpha |
t G_{0} V_{123}^{(3)} |\Psi\rangle  \biggr\}. \,\,\,
\end{eqnarray}
The matrix elements of the second term can be evaluated 
as:
\begin{eqnarray}
\label{Eq.3NF-completeness}  \langle \, {\bf p}\, {\bf q}\, \alpha |
t G_{0} V_{123}^{(3)} |\Psi\rangle &=&
  \sum_{\gamma' , \gamma'', \alpha'''} \, \int d^{3} p''' \int d^{3} q''' \frac{g_{\gamma' \alpha}
\, g_{\gamma'' \alpha'''}}{{E-\frac{p'''^{2}}{m}
-\frac{3q'''^{2}}{4m}}}
 \nonumber \\ &\times& \langle \, {\bf p}\, {\bf q} \, \gamma' |t  |
{\bf p}'''\, {\bf q}''' \, \gamma'' \, \rangle    \langle {\bf
p}'''\, {\bf q}''' \, \alpha'''\, |V_{123}^{(3)} |\Psi\rangle,
\end{eqnarray}
after evaluating the matrix elements of the NN $t$-matrix 
and integrating over ${\bf q}'''$ vector, one obtains:
\begin{eqnarray}
\label{Eq.3NF-t-evaluated}
 \langle \,
{\bf p}\, {\bf q}\, \alpha | t G_{0} V_{123}^{(3)} |\Psi\rangle &=&
    \sum_{\gamma',\gamma'',\alpha'''} \,  \int d^{3} p''' \frac{g_{\gamma' \alpha}
\, g_{\gamma'' \alpha'''}}{{E-\frac{p'''^{2}}{m}
-\frac{3q^{2}}{4m}}} \, \delta_{m'_{s_{3}} m''_{s_{3}}} \,
\delta_{m'_{t_{3}} m''_{t_{3}}} \nonumber
\\ && \hspace{-15mm} \times
 \langle \, {\bf p} \, m'_{s_{1}}
m'_{s_{2}} m'_{t_{1}} m'_{t_{2}} |t(\epsilon) |{\bf p}''' \,
m''_{s_{1}} m''_{s_{2}} m''_{t_{1}} m''_{t_{2}} \rangle \, \langle
{\bf p}'''  \, {\bf q}  \, \alpha'''\, |V_{123}^{(3)} |\Psi\rangle.
\end{eqnarray}
By using the symmetry property of 3NF and the anti-symmetry property
of the total wave function under exchange of nucleons 1 and 2, one
can rewrite Eq. (\ref{Eq.3NF-t-evaluated}) as:
\begin{eqnarray}
\label{Eq.3NF-t-under-p}  \langle \, {\bf p}\, {\bf q}\, \alpha | t
G_{0} V_{123}^{(3)} |\Psi\rangle &=&
    \sum_{\gamma',\gamma'',\alpha'''} \,  \int d^{3} p''' \frac{g_{\gamma' \alpha}
\, g_{\gamma'' \alpha'''} \, \delta_{m'_{s_{3}} m''_{s_{3}}} \,
\delta_{m'_{t_{3}} m''_{t_{3}}} }{{E-\frac{p'''^{2}}{m}
-\frac{3q^{2}}{4m}}} \,  \left(- (-)^{s_{12}'''+t_{12}'''} \right)
\nonumber
\\ && \hspace{-25mm} \times
 \langle \, {\bf p} \, m'_{s_{1}}
m'_{s_{2}} m'_{t_{1}} m'_{t_{2}} |t(\epsilon) P_{12} |-{\bf p}''' \,
m''_{s_{2}} m''_{s_{1}} m''_{t_{2}} m''_{t_{1}} \rangle   \langle
-{\bf p}'''  \, {\bf q} \, \alpha'''\, |V_{123}^{(3)} |\Psi\rangle,
\end{eqnarray}
under the exchange of the labels $m''_{s_{1}}, m''_{t_{1}}$ to
$m''_{s_{2}}, m''_{t_{2}}$, reverse of it and changing ${\bf p}''' $
to $-{\bf p}''' $ one finds:
\begin{eqnarray}
\label{Eq.3NF-exchange-lables}  \langle \, {\bf p}\, {\bf q}\,
\alpha | t G_{0} V_{123}^{(3)} |\Psi\rangle &=&
    \sum_{\gamma',\gamma'',\alpha'''} \,   \int d^{3} p''' \frac{g_{\gamma' \alpha}
\, g_{\gamma'' \alpha'''}}{{E-\frac{p'''^{2}}{m}
-\frac{3q^{2}}{4m}}} \,\, \delta_{m'_{s_{3}} m''_{s_{3}}} \,
\delta_{m'_{t_{3}} m''_{t_{3}}} \nonumber
\\ && \hspace{-25mm} \times
 \langle \, {\bf p} \, m'_{s_{1}}
m'_{s_{2}} m'_{t_{1}} m'_{t_{2}} |-t(\epsilon) P_{12} |{\bf p}''' \,
m''_{s_{1}} m''_{s_{2}} m''_{t_{1}} m''_{t_{2}} \rangle \, \langle
{\bf p}'''  \, {\bf q}  \, \alpha'''\, |V_{123}^{(3)} |\Psi\rangle.
\end{eqnarray}
Now one can consider Eqs. (\ref{Eq.3NF-t-evaluated}) and
 (\ref{Eq.3NF-exchange-lables}) together to achieve:
\begin{eqnarray}
\label{Eq.3NF-add} \langle \, {\bf p}\, {\bf q}\, \alpha | t G_{0}
V_{123}^{(3)} |\Psi\rangle &=& \frac{1}{2}
    \sum_{\gamma',\gamma'',\alpha'''} \,   \int d^{3} p''' \frac{g_{\gamma' \alpha}
\, g_{\gamma'' \alpha'''}}{{E-\frac{p'''^{2}}{m}
-\frac{3q^{2}}{4m}}} \,\, \delta_{m'_{s_{3}} m''_{s_{3}}} \,
\delta_{m'_{t_{3}} m''_{t_{3}}} \nonumber
\\ && \hspace{-30mm} \times  \langle \, {\bf p} \, m'_{s_{1}}
m'_{s_{2}} m'_{t_{1}} m'_{t_{2}} |t(\epsilon) (1-P_{12}) |{\bf p}'''
\, m''_{s_{1}} m''_{s_{2}} m''_{t_{1}} m''_{t_{2}} \rangle \,
\langle {\bf p}'''  \, {\bf q}  \, \alpha'''\, |V_{123}^{(3)}
|\Psi\rangle,
\end{eqnarray}
by applying the introduction of the anti-symmetrized NN $t$-matrix 
one can obtain the coupled three-dimensional Faddeev integral
equations as:
\begin{eqnarray}
\label{FCs} \langle \, {\bf p}\, {\bf q}\, \alpha \, |\psi\rangle
&=& \frac{1}{{E-\frac{p^{2}}{m} -\frac{3q^{2}}{4m}}} \nonumber
\\ && \hspace{-20mm} \times \Biggl [\,
  \int d^{3}q' \, \sum_{\gamma',\gamma''',\alpha''} \,
g_{\alpha \gamma'''} \, g_{\gamma' \alpha''} \, \delta_{m'''_{s_{3}}
m'_{s_{1}}} \, \delta_{m'''_{t_{3}} m'_{t_{1}}} \nonumber
\\  && \hspace{-18mm} \quad \times \,\, _{a}\langle{\bf p}\, m'''_{s_{1}}
m'''_{s_{2}} \, m'''_{t_{1}} m'''_{t_{2}} |t(\epsilon)
|\frac{-1}{2}{\bf q} -{\bf q}' \, m'_{s_{2}} m'_{s_{3}} \,
m'_{t_{2}} m'_{t_{3}} \rangle_{a} \,  \langle{\bf q}+\frac{1}{2}
{\bf q}' \,\, {\bf q}' \, \alpha''|\psi\rangle
 \nonumber \\  && \hspace{-20mm} \quad +
 \Biggl\{\,\, \langle \, {\bf p}\, {\bf q}\, \alpha \, |V_{123}^{(3)}
|\Psi\rangle \nonumber \\  && \hspace{-10mm} + \frac{1}{2}
\sum_{\gamma',\gamma'',\alpha'''} \,  g_{\alpha \gamma'} \,
g_{\gamma'' \alpha'''}  \int d^{3}p'\, \frac{\delta_{m'_{s_{3}}
m''_{s_{3}}} \delta_{m'_{t_{3}} m''_{t_{3}}} }{E-\frac{p'^{2}}{m}
-\frac{3q^{2}}{4m} } \nonumber \\ && \hspace{-10mm} \times \,
_{a}\langle{\bf p}\, m'_{s_{1}} m'_{s_{2}} \, m'_{t_{1}} m'_{t_{2}}
|t(\epsilon) |{\bf p}' \, m''_{s_{1}} m''_{s_{2}} \, m''_{t_{1}}
m''_{t_{1}} \rangle_{a} \, \langle{\bf p}'\,{\bf q} \,
\alpha'''|V_{123}^{(3)} |\Psi\rangle \Biggr\} \, \, \Biggr].
\end{eqnarray}
To represent the generality of our 3D formalism we can simplify the
Eq. (\ref{FCs}) to the bosonic case by switching off the
spin-isospin quantum numbers, see Ref. \cite{Liu-FBS33}. 
%
In order to show the efficiency of the presented formalism the
realistic $2\pi$-exchange TM 3NF \cite{Coon-PRC23} has been used to
evaluate the matrix elements of $\langle \, {\bf p}\, {\bf q}\,
\alpha \, |V_{123}^{(3)} |\Psi\rangle$.

\section{The Evaluation of $\langle \, {\bf p}\, {\bf q}\, \alpha \, |V_{123}^{(3)} |\Psi\rangle$ for the TM 3NF }
\label{appendix:TM} For evaluation of the coupled three dimensional
Faddeev equations, the matrix elements $\langle \, {\bf p}\, {\bf
q}\, \alpha \, |V_{123}^{(3)} |\Psi\rangle$ need to be calculated.
In this section these matrix elements have been evaluated for the TM
$2\pi$-exchange 3NF. To this aim one should first prepare this force
in the form that can be evaluated easily in 3D representation as:
\begin{eqnarray}
\label{TM}
 \nonumber V_{123}^{(3)}&=& V_{0} Q \ Q' \ \sigma_{1}\cdot \widehat{{\bf Q}} \, \sigma_{2}\cdot
\widehat{{\bf Q}}'  \,  {F({Q}^{2})\over  {Q}^{2}+m_{\pi}^{2}}\
{F({Q}'^{2})\over {Q}'^{2}+m_{\pi}^{2}}\\
&\times& \Biggl[ \tau_{1}\cdot \tau_{2}\ \biggl( A+B\  Q Q' \gamma
+C\ ({Q}^{2}+{Q}'^{2})\biggl) \nonumber \\ && \,\, +D\ \tau_{3}\cdot
\tau_{1}\times\tau_{2}\ \biggl \{ \frac{3ia}{4\gamma} -\frac{i}{2
\gamma a} \ (\sigma_{3}\cdot \widehat{{\bf
 Q}}')^{2}+\frac{i}{2 a} \ \sigma_{3}\cdot \widehat{{\bf
 Q}}' \ \sigma_{3}\cdot \widehat{{\bf Q}} \nonumber \\ && \hspace{+28mm} +\frac{i}{8 \gamma a} \ (\sigma_{3}\cdot \widehat{{\bf
 Q}}')^{2} \ (\sigma_{3}\cdot \widehat{{\bf Q}})^{2} - \frac{i}{2 \gamma a} \ (\sigma_{3}\cdot \widehat{{\bf
 Q}})^{2}  \biggr \}
\Biggr],
\end{eqnarray}
where $\gamma= \widehat{{\bf Q}}.\widehat{{\bf Q}}'$ and
$a=\sqrt{1-\gamma^{2}}$. The ${\bf Q}$ and ${\bf Q}'$ are momentum
transfers, $\sigma_i$'s are Pauli spin matrices and $ F(Q^{2})$ are
form factors.
One distinguishes four $A$-, $B$-, $C$- and $D$-terms in the TM
force. The scalar product of the spin-momentum vectors can be
evaluated as:
\begin{eqnarray}
\label{s_dot_Q} \langle \, \hat{z} \, m'_{s} | \sigma \cdot
\widehat{{\bf Q}} | \hat{z}\, m''_{s} \, \rangle \ = \sum_{m_{s}}
m_{s} \ D_{m'_{s} m_{s}}^{\frac{1}{2}} (\widehat{{\bf Q}}) \ \
D_{m''_{s} m_{s}}^{\frac{1}{2}\,*} (\widehat{{\bf Q}}) = {\O}
_{m'_{s} m''_{s}} ^{ \widehat{{\bf Q}}},
\end{eqnarray}
where $D_{m'_{s} m_{s}}^{\frac{1}{2}}$ is Wigner D-function which is
defined generally as $D_{m'_{s} m_{s}}^{s} (\hat{q})= \langle
\hat{z} s m'_{s} | \hat{q} s m_{s} \rangle$. The application of the
TM 3NF to the total wave function $|\Psi\rangle$ can be considered
as sum of the the four independent terms:
\begin{eqnarray}
  \label{psi}
 V_{123}^{(3)} |\Psi\rangle= \sum_{i=A}^{D} V_{31}^{(i)} \ I^{(i)}
\ V_{23}^{(i)} | \Psi \rangle
 = \sum_{i=A}^{D}  | \psi^{i} \rangle, \quad I^{(i)}=
  \left\{%
    \begin{array}{ll}
    \tau_{1}\cdot \tau_{2}\, & \hbox{i=A, B, C} \\
    \tau_{3}\cdot \tau_{1}\times\tau_{2}\, & \hbox{i=D}. \\
    \end{array}%
\right.
 \end{eqnarray}
In the following the matrix elements of $\langle \, {\bf p}\, {\bf
q}\, \alpha \, |\psi^{i}\rangle$ terms have been evaluated. It is
clear that $V_{123}^{(3)}$ can be splitted into two parts and each
part contains a meson exchange, the mesons are exchanged in the
subsystems (31) and (23), which are called for convenience subsystem
2 and 1 correspondingly. So it is convenient to insert a complete
set of states of type $2$ between $V_{123}^{(3)}$ and $|\Psi\rangle$
and another complete set of states of type $1$ between the two meson
exchanges. Then the matrix elements of $\psi^{i}$ can be written as:
\begin{eqnarray}
\label{psi_u} _{3}\langle \, {\bf p}\, {\bf q}\, \alpha \, |\psi^{i}
\rangle &=& \sum_{\alpha'}  \int d^3 p'  \int d^3 q'  \,\,
_{3}\langle \, {\bf p}\, {\bf q}\,
\alpha |  \, {\bf p}'\, {\bf q}' \, \alpha' \, \rangle_{1} \nonumber \\
&\times& \sum_{\alpha''}  \int d^3 p''  \int d^3 q''  \,\,
_{1}\langle \, {\bf p}' \, {\bf q}'\, \alpha' |V_{31}^{(i)}| {\bf
p}'' \, {\bf q}''\, \alpha'' \, \rangle_{1} \nonumber
\\ &\times&
 \sum_{\alpha'''}  \int d^3 p'''  \int d^3 q'''  \,\, _{1}\langle \, {\bf p}'' \, {\bf q}''\, \alpha''
 |I^{(i)}|
 \, {\bf p}''' \, {\bf q}'''\, \alpha''' \, \rangle_{2}
 \nonumber \\ &\times& \sum_{\alpha''''}  \int d^3 p''''  \int d^3 q''''  \,\,
_{2}\langle \, {\bf p}''' \, {\bf q}'''\, \alpha''' \,
|V_{23}^{(i)}| {\bf p}'''' \, {\bf q}''''\, \alpha'''' \,\rangle_{2}  \nonumber \\
&\times& \,_{2} \langle \, {\bf p}'''' \, {\bf q}''''\, \alpha''''
|\Psi\rangle,
\end{eqnarray}
Here the subscripts $1, 2, 3$ of the bra and ket vectors stand for
different 3N chains. The coordinate transformation from the system
of type 1 to one of type 3 can be evaluated explicitly as:
\begin{eqnarray}
\label{u_1_to_3}
 _{3}\langle \, {\bf p}\, {\bf q}\, \alpha
|  \, {\bf p}' \, {\bf q}'\, \alpha' \, \rangle_{1} =g_{\alpha_{3}
\alpha'_{1}} \, \delta^{3}({\bf p}'+\frac{1}{2}{\bf
p}+\frac{3}{4}{\bf q}) \delta^{3}({\bf q}'-{\bf p}+\frac{1}{2}{\bf
q}) .
\end{eqnarray}
The matrix elements of the isospin coordinate transformations can be
evaluated as:
\begin{eqnarray}
\label{I_u}   _{1}\langle \, {\bf p}'' \, {\bf q}''\, \alpha''
 |I^{(i)}|
 \, {\bf p}''' \, {\bf q}'''\, \alpha''' \, \rangle_{2} = \ _{1}\langle \, {\bf p}'' \, {\bf q}''\, \alpha''_{S}
 | {\bf p}''' \, {\bf q}'''\, \alpha'''_{S} \, \rangle_{2} \ \  _{1}\langle \, \alpha''_{T}
 |I^{(i)}|
 \, \alpha'''_{T} \, \rangle_{2},
\end{eqnarray}
\begin{eqnarray}
\label{u_2_to_1}
  _{1}\langle \, {\bf p}'' \, {\bf q}''\, \alpha''_{S}
 | {\bf p}''' \, {\bf q}'''\, \alpha'''_{S} \, \rangle_{2} \ = g_{\alpha''_{1}
 \alpha'''_{2}}^{S} \,
 \delta^{3}({\bf p}''' +\frac{1}{2}
 {\bf p}'' +\frac{3}{4}{\bf q}'' ) \delta^{3}({\bf q}''' -{\bf p}''+\frac{1}{2}{\bf
 q}'').
\end{eqnarray}
The matrix elements of 
$_{1}\langle \, \alpha''_{T} |I^{(i)}|  \, \alpha'''_{T} \,
\rangle_{2}$
 have been derived in Ref. \cite{Huber-FBS22}.

 In the following the matrix elements for the different $V$'s
are evaluated. In the first step the matrix elements of $V_{31}^{i}$
are evaluated. Since both pion-exchange propagators in the 3NF term
depend only on the momentum transfer in a two-nucleon subsystem and
also because of the separation of the isospin parts of 3NF, the
matrix elements of $V_{31}^{i}$ can be evaluated as:
\begin{eqnarray}
\label{W_31^i} _{1}\langle \, {\bf p}' \, {\bf q}'\, \alpha'
|V_{31}^{(i)}| {\bf p}'' \, {\bf q}''\, \alpha'' \, \rangle_{1} =
\delta^{3}({\bf p}'-{\bf p}'') \  \delta_{\alpha'_{T} \alpha''_{T}}
\ _{1}\langle \, {\bf q}'\, \alpha'_{S} |V_{31}^{(i)}| {\bf q}''\,
\alpha''_{S} \, \rangle_{1},
\end{eqnarray}
and the spin-space parts can be more simplified for the
$A$-,$B$-,$C$- and $D$-terms separately as:
\begin{eqnarray}
\label{W_31^ABC}
 _{1}\langle \, {\bf q}'\, \alpha'_{S}
|V_{31}^{(A, B, C)}| {\bf q}''\, \alpha''_{S} \, \rangle_{1} &=&
\sum_{\gamma'_{S},\gamma''_{S}} g_{\alpha'_{S} \gamma'_{S}}^{1} \
g_{\alpha''_{S} \gamma''_{S}}^{1} \ \delta_{m'_{s_{2}} m''_{s_{2}}}
\delta_{m'_{s_{3}} m''_{s_{3}}} \nonumber \\ &\times& \ _{1}\langle
\, {\bf q}'\, m'_{s_{1}} |V_{31}^{(A, B, C)}| {\bf q}''\,
m''_{s_{1}} \, \rangle_{1}, \nonumber \\
_{1}\langle \, {\bf q}'\, \alpha'_{S} |V_{31}^{(D)}| {\bf q}''\,
\alpha''_{S} \, \rangle_{1} &=& \sum_{\gamma'_{S}, \gamma''_{S}}
g_{\alpha'_{S} \gamma'_{S}}^{1} \ g_{\alpha''_{S} \gamma''_{S}}^{1}
\ \delta_{m'_{s_{2}} m''_{s_{2}}} \nonumber
\\ &\times& \ _{1}\langle \, {\bf q}'\, m'_{s_{1}} \, m'_{s_{3}}
|V_{31}^{(D)}| {\bf q}''\, m''_{s_{1}} \, m''_{s_{3}} \,
\rangle_{1}.
\end{eqnarray}
The $A$-, $B$- and $C$-terms can be evaluated as:
\begin{eqnarray}
\label{W_31^A_B_C}
 _{1}\langle \, {\bf q}'\, m'_{s_{1}} |V_{31}^{(A,B,C)} | {\bf q}''\, m''_{s_{1}} \, \rangle_{1} &=& \frac{F(({\bf q}'-
 {\bf q}'')^{2})}{({\bf q}'-{\bf q}'')^{2}+m_{\pi}^{2}} \ |{\bf
q}'-{\bf q}''|^N \ {\O} _{m'_{s_{1}} m''_{s_{1}}} ^{ \widehat{({\bf
q}'-{\bf q}'')} },
\end{eqnarray}
where the value of N is 1, 2 and 3 corresponding to $A$-, $B$- and
$C$-terms. For the different parts, five parts, of the $D$-term one
obtains:
\begin{eqnarray}
\label{W_31^D_1-5} _{1}\langle \, {\bf q}'\, m'_{s_{1}} \,
m'_{s_{3}} |V_{31}^{(D)} | {\bf q}''\, m''_{s_{1}} \, m''_{s_{3}} \,
\rangle_{1}  &=&  \, _{1}\langle \, {\bf q}'\, m'_{s_{1}}
|V_{31}^{(B)} | {\bf q}''\, m''_{s_{1}} \, \rangle_{1} \ {\O}
_{m'_{s_{3}} m''_{s_{3}}} ^{ \widehat{({\bf q}'-{\bf q}'')} } \
^{M},
\end{eqnarray}
where the value of M is 0 for the first and second parts of this
term, it is 1 for the third part and it is 2 for the fourth and
fifth parts. The matrix elements of $V_{23}^{i}$ can be evaluated by
following the same algorithm as above. So one can obtain:
\begin{eqnarray}
\label{W_23^i}  _{2}\langle \, {\bf p}''' \, {\bf q}'''\, \alpha'''
\, |V_{23}^{(i)}| {\bf p}'''' \, {\bf q}''''\, \alpha''''
\,\rangle_{2} &=&
 \delta^{3}({\bf p}''' -{\bf p}'''') \  \delta_{\alpha'''_{T} \alpha''''_{T}} \
 _{2}\langle \, {\bf q}'''\, \alpha'''_{S}
\, |V_{23}^{(i)}| {\bf q}''''\, \alpha''''_{S} \,\rangle_{2},
\nonumber \\
\end{eqnarray}
\begin{eqnarray}
\label{W_23^ABC} _{2}\langle \, {\bf q}'''\, \alpha'''_{S}
|V_{23}^{(A, B, C)}| {\bf q}''''\, \alpha''''_{S} \, \rangle_{2} &=&
\sum_{\gamma'''_{S}, \gamma''''_{S}} g_{\alpha'''_{S}
\gamma'''_{S}}^{2} \ g_{\alpha''''_{S} \gamma''''_{S}}^{2} \
\delta_{m'''_{s_{1}} m''''_{s_{1}}} \delta_{m'''_{s_{3}}
m''''_{s_{3}}}  \nonumber \\
&\times& \ _{2}\langle \, {\bf q}'''\, m'''_{s_{2}} |V_{23}^{(A, B,
C)}| {\bf q}''''\, m''''_{s_{2}} \, \rangle_{2},
\end{eqnarray}
\begin{eqnarray}
\label{W_23^D} _{2}\langle \, {\bf q}'''\, \alpha'''_{S}
|V_{23}^{(D)}| {\bf q}''''\, \alpha''''_{S} \, \rangle_{2} &=&
\sum_{\gamma'''_{S}, \gamma''''_{S}}  g_{\alpha'''_{S}
\gamma'''_{S}}^{2} \ g_{\alpha''''_{S} \gamma''''_{S}}^{2} \
\delta_{m'''_{s_{1}} m''''_{s_{1}}}
 \nonumber \\
&\times& \ _{2}\langle \, {\bf q}'''\, m'''_{s_{2}} \, m'''_{s_{3}}
|V_{23}^{(D)}| {\bf q}''''\, m''''_{s_{2}} \, m''''_{s_{3}} \,
\rangle_{2}.
\end{eqnarray}

The $A$-, $B$- and $C$-terms which have been considered in the right
side of Eq. (\ref{W_23^ABC}) can be evaluated as:
\begin{eqnarray}
\label{W_31^A_B_C}
 _{2}\langle \, {\bf q}'''\, m'''_{s_{2}} |V_{23}^{(A,B,C)} | {\bf q}''''\, m''''_{s_{2}} \, \rangle_{2} =  \frac{F(({\bf q}''''-{\bf
q}''')^{2})}{({\bf q}''''-{\bf q}''')^{2}+m_{\pi}^{2}} \ |{\bf
q}''''-{\bf q}'''| ^N \ {\O} _{m'''_{s_{2}} m''''_{s_{2}}} ^{
\widehat{({\bf q}''''-{\bf q}''')} }, \,\,\,\,
\end{eqnarray}
where the value of N is 1, 2 and 3 corresponding to $A$-, $B$- and
$C$-terms. Also the different parts of the $D$-term which have been
considered in the right side of Eq. (\ref{W_23^D}) can be evaluated
in the following:
\begin{eqnarray}
\label{W_31^D_1-5}  _{2}\langle \, {\bf q}'''\, m'''_{s_{2}} \,
m'''_{s_{3}} |V_{23}^{(D)} | {\bf q}''''\, m''''_{s_{1}} \,
m''''_{s_{3}} \, \rangle_{2}  =  \, _{2}\langle \, {\bf q}'''\,
m'''_{s_{2}} |V_{23}^{(B)} | {\bf q}''''\, m''''_{s_{1}} \,
\rangle_{2} \ {\O} _{m'''_{s_{2}} m''''_{s_{2}}} ^{ \widehat{({\bf
q}''''-{\bf q}''')} } \ ^{M}, \,\,\,\,\,\,
\end{eqnarray}
where the value of M is 0 for the first and fifth parts of this
term, it is 1 for the third part and it is 2 for the second and
fourth parts.

\section{Summary}\label{sec:summary}

A new formalism of three dimensional Faddeev integral equations for
the 3N bound state including the 3NF is proposed. This formulation
leads to a coupled set of a strictly finite number of equations in
two vector variables for the amplitudes. The comparison of 3D and PW
formalisms shows that this non PW formalism avoids the very involved
angular momentum algebra occurring for the permutations and
transformations and it is more efficient and less cumbersome for
considering the 3NF. This formalism enables us to handle the
realistic 2N and 3N forces with all their complexity in 3N bound
state calculations.

\section*{Acknowledgments}
M. R. H. would like to thank the organizers of the APFB08 conference
for their kind invitation, warm hospitality and support, and the
very pleasant and stimulating atmosphere at the conference. We would
like to express our acknowledgements to our colleagues M. Harzchi
and M. A. Shalchi for the enriching discussions. This work was
supported by the research council of the University of Tehran.

\end{document}